\documentclass[twocolumn,showpacs]{revtex4}

\usepackage{graphicx}% Include figure files
\usepackage{dcolumn}% Align table columns on decimal point
\usepackage{bm}% bold math
\usepackage{amssymb}

\begin{document}

\preprint{}

\title{Lorentz Violation of The Standard Model}% Force line breaks with \\

\author{J.C. Yoon}
\email{jcyoon@u.washington.edu} \affiliation{University of
Washington}

\date{\today}% It is always \today, today,
             %  but any date may be explicitly specified

\begin{abstract}
Though the standard model of electroweak
interactions\cite{Glashow,Weinberg,Salam} is commonly believed to
provide a successful unification of electromagnetic and weak
interactions, the approximation in the massless limit and the
assumption of massless fermion in this model should be
investigated further with rigorousness. Here it will be shown that
this approximation violates Lorentz invariance and it is still
necessary even with the assumption of a massless fermion and the
Higgs mechanism\cite{Higgs,Kibble}. We conclude that the
unification of electroweak interactions is only valid with the
assumption of Lorentz violation.
\end{abstract}

% PACS, the Physics and Astronomy
                             % Classification Scheme.
%\keywords{Lorentz violation, Standard Model}%Use showkeys class option if keyword
                              %display desired
\maketitle
\section{Introduction}
The standard model of the Glashow-Weinberg-Salam theory, suggested
as a unified model of electroweak interactions, is based on the
massless approximation and the assumption of massless fermion, not
just that of massless neutrino to be accurate. Here the physical
argument of electroweak unification will be investigated after
clarifying the ambiguity in the helicity and chirality of fermion
following the standard textbook\cite{PeskinSchroeder}

\section{Helicity and Chirality}
A Dirac field $\psi$ can be written as a linear combination of
plane waves since it obeys the Klein-Gordon equation.
\begin{eqnarray}
\psi(x) = u(p)e^{-ip \cdot x} \nonumber
\end{eqnarray}
where $p^2 = m^2$ and let us consider solutions with positive
frequency, $p^0 > 0$, for simplicity. One may obtain the solutions
of the Dirac equation in the rest frame, $p=p_{0}=(m,\mathrm{0})$
\begin{eqnarray}
u(p_{0}) = \sqrt{m} \left(
\begin{array}{c} \xi \\ \xi
\end{array} \right) \nonumber
\end{eqnarray}
for any numerical two-component spinor $\xi$. The general form of
$u(p)$ in other frame can be derived by boosting $u(p_{0})$ in the
rest frame. Let us define $\eta$, {\it rapidity} by considering a
boost along the $z$-direction to the 4-momentum vector. For finite
$\eta$,
\begin{eqnarray}
\left( \begin{array}{c}  E \\ p^{3} \end{array} \right) & = & \exp
\bigg[ \eta \left( \begin{array}{cc}  0 & 1 \\ 1 & 0 \end{array}
\right) \bigg] \left( \begin{array}{c}  m \\ 0 \end{array} \right) \nonumber \\
&=&  \left( \begin{array}{c}  m\cosh \eta \\ m \sinh \eta
\end{array} \right) \nonumber
\end{eqnarray}

Applying the same boost to $u(p_{0})$,
\begin{eqnarray}
u(p) &=& \exp \bigg[ - {1 \over 2} \eta \left( \begin{array}{cc}
\sigma^{3} & 0 \\ 0 & -\sigma^{3} \end{array} \right)   \bigg]
\sqrt{m} \left( \begin{array}{c} \xi \\ \xi \end{array} \right)
\nonumber \\
&=& \left(
\begin{array}{c} \big[ \sqrt{E+p^{3}}({ 1-\sigma^{3} \over 2 }) +
 \sqrt{E-p^{3}}({ 1+\sigma^{3} \over 2 }) \big] \xi \\
 \big[\sqrt{E+p^{3}}({ 1+\sigma^{3} \over 2 }) +
 \sqrt{E-p^{3}}({ 1-\sigma^{3} \over 2 })
 \big] \xi
\end{array} \right) \nonumber
\end{eqnarray}
If $\xi = \left( \begin{array}{c} 1 \\
0\end{array} \right)$(spin up along the $z$-axis) in the boosted
frame of $z$ direction,
\begin{eqnarray}
u_{z+;h+} = \left(
\begin{array}{c} \sqrt{E - p^{3}}
         \left( \begin{array}{c} 1 \\ 0  \end{array} \right) \\
\sqrt{E + p^{3}}
         \left( \begin{array}{c} 1 \\ 0 \end{array} \right)
\end{array} \right) \nonumber
\end{eqnarray}
while for $\xi = \left( \begin{array}{c} 1 \\ 0 \end{array}
\right)$ (spin down along the $z$-axis) in the boosted frame of
$z$ direction we have
\begin{eqnarray}
u_{z+;h-} = \left(
\begin{array}{c} \sqrt{E + p^{3}}
         \left( \begin{array}{c} 0 \\ 1  \end{array} \right) \\
\sqrt{E - p^{3}}
         \left( \begin{array}{c} 0 \\ 1 \end{array} \right)
\end{array} \right) \nonumber
\end{eqnarray}
Following the same steps, in the boosted frame of $-z$ direction
we have
\begin{eqnarray}
u_{z-;h+} = \left(
\begin{array}{c} \sqrt{E - p^{3}}
         \left( \begin{array}{c} 0 \\ 1  \end{array} \right) \\
\sqrt{E + p^{3}}
         \left( \begin{array}{c} 0 \\ 1 \end{array} \right)
\end{array} \right) \nonumber \\
u_{z-;h-} = \left(
\begin{array}{c} \sqrt{E + p^{3}}
         \left( \begin{array}{c} 1 \\ 0  \end{array} \right) \\
\sqrt{E - p^{3}}
         \left( \begin{array}{c} 1 \\ 0 \end{array} \right)
\end{array} \right) \nonumber
\end{eqnarray}
where ${\it helicity}$ $h \pm$ is defined by the momentum of
fermion and its spin orientation: if spin orientation is in the
same direction as its momentum, it is called right-handed
helicity($h+$). Therefore, a massive fermion field of the Dirac
equation can be described by two nonzero Weyl spinors with
left-handed and right-handed chirality
\begin{eqnarray}
\Psi  = \left( \begin{array}{c}  \psi_{L} \\ \psi_{R} \end{array}
\right)  \nonumber
\end{eqnarray}
where $\psi_{L,R} \neq 0$ for a massive fermion and ${\it
chirality}$ $(L,R)$ is defined to indicate either of these
two-component objects. However, the chirality is not a physical
observable unlike the helicity since no corresponding physical
measurement is available and also a massive fermion satisfying the
Dirac equation cannot be represented by only one chirality since
it only denotes a part of the solution to the Dirac equations are
there are always two chiralities for a massive
particle($\psi_{L,R} \neq 0$). The helicity of a massive particle
is also not well defined in the rest frame where its momentum is
zero and the left-handed and right-handed particle are
indistinguishable. It is also not a fundamental property of
particle since it is not conserved under Lorentz transformations:
one can always find a boosted frame, where the helicity is
opposite due to its reversed momentum, for example, $u_{z+;h+}$
can be observed as $u_{z-;h-}$ in a boosted reference frame.
Therefore, the helicity of a massive fermion is not a
Lorentz-invariant observable and it cannot be considered as a
fundamental property of particle.

A massless fermion can be derived by taking the massless limit or
solving massless Dirac equation. For the boosted frame of $z$
direction we have
\begin{eqnarray}
u_{z+;h+} = \sqrt{E + p^{3}} \left(
\begin{array}{c}
          0 \\ 0   \\
          1 \\ 0
\end{array} \right)
u_{z+;h-} = \sqrt{E + p^{3}} \left(
\begin{array}{c}
          0 \\ 1  \\
          0 \\ 0
\end{array} \right) \nonumber
\end{eqnarray}
In the boosted frame of $-z$ direction,
\begin{eqnarray}
u_{z-;h+} = \sqrt{E + p^{3}} \left(
\begin{array}{c}
          0 \\ 0   \\
          0 \\ 1
\end{array} \right)
u_{z-;h-} = \sqrt{E + p^{3}} \left(
\begin{array}{c}
          1 \\ 0   \\
          0 \\ 0
\end{array} \right) \nonumber
\end{eqnarray}
Unlike massive fermions, they can be represented by only one
chirality
\begin{eqnarray}
\Psi  = \left( \begin{array}{c}  \psi_{L} \\ 0 \end{array} \right)
\mathrm{or} \left( \begin{array}{c}  0 \\ \psi_{R}
\end{array} \right) \nonumber
\end{eqnarray}
Since a massless fermion has neither the rest frame nor a boosted
frame which reverses its helicity, the helicity is well defined
and can be considered as a fundamental property of particle that
is Lorentz invariant. Its chirality also can be considered as
fundamental property as it corresponds to its helicity, which
provides the physical measurement of chirality.

The distinction of massive and massless particles in chirality and
helicity should be treated with care in approximating a massive
fermion in the massless limit. A massive fermion in a large boost
can be approximately described as a massless particle. However,
the exact solutions of the massive Dirac equation in the rest
frame cannot be achieved by Lorentz transformations of the
approximated ones back to the rest frame, for example,
\begin{eqnarray}
u &=& \sqrt{m} \left(
\begin{array}{c} \xi \\ \xi
\end{array} \right) \qquad \qquad {in~the~rest~frame} \nonumber \\
& \rightarrow &  \left(
\begin{array}{c} \sqrt{E + p^{3}} \xi \\
\sqrt{E - p^{3}} \xi
\end{array} \right)  \qquad {in~the~boosted~frame} \nonumber \\
&\rightarrow & \sqrt{2E} \left( \begin{array}{c}  \xi \\ 0
\end{array} \right) \qquad \qquad {approximation} \nonumber \\
&\nrightarrow& \sqrt{m} \left(
\begin{array}{c} \xi \\ \xi
\end{array} \right) \qquad \qquad{back~to~the~rest~frame} \nonumber
\end{eqnarray}
Therefore, the massless approximation violates Lorentz invariance
and thus it should not be justified to determine the fundamental
property of a massive particle as that of massless one.

\section{The Standard Model}
The weak interactions have a certain structure of ($1 \pm
\gamma^5$)\cite{FeynmanGellMann,Sudarshan} as
\begin{eqnarray}
H_{weak} = \overline{\psi} \gamma^{\mu} {(1 \pm \gamma^5)\over 2}
\psi, \nonumber
\end{eqnarray}
while the electromagnetic interactions are
\begin{eqnarray}
H_{em} = \overline{\psi} \gamma^{\mu} \psi \nonumber
\end{eqnarray}
In the standard model, they are unified as the electroweak
interactions in
\begin{eqnarray}
H_{ew} = \overline{\psi}_{L,R} \gamma^{\mu} \psi_{L,R} \nonumber
\end{eqnarray}
as massive fermions are represented with definite helicities in
the massless limit\cite{Glashow,Weinberg,Salam}. This model
suggests to interpret $ {1 \over 2} (1 \pm \gamma^5)$ not as a
structure of interactions, but as a physical operator acting on
fermion fields so that the weak interactions are
\begin{eqnarray}
H_{weak} &=& \overline{\psi} \gamma^{\mu} {(1 \pm \gamma^5)\over
2}
\psi \nonumber \\
&=& \overline{\psi} \gamma^{\mu} \left[{(1 \pm \gamma^5)\over
2}\right]^{2}
\psi \nonumber \\
&=& \overline{\psi} {(1 \mp \gamma^5)\over 2} \gamma^{\mu} {(1 \pm
\gamma^5)\over 2}
\psi \nonumber \\
&=& \overline{\psi}_{L,R} \gamma^{\mu} \psi_{L,R} \nonumber
\end{eqnarray}
where
\begin{eqnarray}
\psi_{L}  &=& {1 \over 2 }(1 - \gamma^{5})\psi \nonumber
 = \left( \begin{array}{c}  \psi'_{L} \\ 0 \end{array}
\right) \nonumber \\
\psi_{R}  &=&  {1 \over 2 }(1 + \gamma^{5})\psi \nonumber = \left(
\begin{array}{c}  0 \\ \psi'_{R}
\end{array} \right) \nonumber
\end{eqnarray}
However, no massive fermion of the Dirac equation corresponds to
$\psi_{L,R}$ and thus it is only valid for a massive particle in
the massless limit. We may argue that the calculation of
$H_{weak}$ would be the same regardless of this approximation
because ${1 \over 2} (1 \pm \gamma^5)$ associates with only one
chiral component.
\begin{eqnarray}
H_{weak} &=& \overline{\psi} \gamma^{\mu} {(1 \pm \gamma^5)\over
2}
\psi \nonumber \\
&=& \overline{\psi}_{h \pm} \gamma^{\mu} {(1 \pm \gamma^5)\over 2}
\psi_{h \pm} \nonumber \\
&\rightarrow& \overline{\psi}_{L,R} \gamma^{\mu} {(1 \pm
\gamma^5)\over 2} \psi_{L,R} \nonumber \\
&=& \overline{\psi}_{L,R} \gamma^{\mu} \psi_{L,R} \nonumber
\end{eqnarray}

However, we have two distinctive structure ${1 \over 2} (1 \pm
\gamma^5)$ for a fermion field since whether the particle is left-
or right-handed in some reference frames should not determine
which of these two structures we have in the rest frame and thus
the massless approximation is also required for this argument.
\begin{eqnarray}
{1 \over 2} (1 - \gamma^5)\psi_{h +} &\neq& {1 \over 2} (1 -
\gamma^5)\psi_{R} = 0 \nonumber
\end{eqnarray} while
\begin{eqnarray}
{1 \over 2} (1 + \gamma^5)\psi_{h +} = {1 \over 2} (1 +
\gamma^5)\psi_{R} \nonumber
\end{eqnarray}
The electromagnetic interactions are also approximated since a
massive Dirac fermion that we observe is not $\psi_{L,R}$, but
$\psi_{h \pm}$.
\begin{eqnarray}
H_{em} &=& \overline{\psi} \gamma^{\mu} \psi \nonumber \\
&=& \overline{\psi}_{h\pm} \gamma^{\mu} \psi_{h\pm} \nonumber \\
&\rightarrow& \overline{\psi}_{L,R} \gamma^{\mu} \psi_{L,R}
\nonumber
\end{eqnarray}
Therefore, the unification of electroweak interactions is based on
the approximation that violates Lorentz invariance.

The most accurate description of electroweak interactions should
be obtained in the rest frame where the exact solutions of the
Dirac equations are available and always the same whether they are
observed as left- or right-handed in other reference frames since
the interactions should be Lorentz invariant. In the rest frame,
there are still two different electroweak interactions.
\begin{eqnarray}
H_{ew} &=& \overline{\psi}_{0} \gamma^{\mu} {(1 \pm \gamma^5)\over
2}
\psi_{0} \nonumber \\
&\neq& \overline{\psi}_{0} \gamma^{\mu} \psi_{0}  \nonumber
\end{eqnarray}
Therefore, the unification of electroweak interactions fails in
the most accurate description of interactions and it is only valid
with the assumption of Lorentz violation.

\section{The Higgs Mechanism}
This Lorentz violation of the standard model might be avoided by
the assumption that a fermion is massless as its mass is obtained
from the spontaneous symmetry breaking\cite{Higgs,Kibble}. In the
Higgs mechanism, a fermion is defined as a massless particle with
a definite chirality $\psi_{L} = {1 \over 2 }(1 - \gamma^{5})\psi
$. However, it would be inconsistent if this definition of
massless fermion remains valid even after its mass is acquired
since a massive fermion satisfying the Dirac equation cannot be
represented by only one chirality. When we describe the physics of
a massive fermion, its nonzero mass should be accepted as a given
physical property of particle in the fundamental equations that is
independent of any proceeding physical events that occurred
before: a fermion field should be defined by the massive Dirac
equation regardless of how its mass is given. Since the Higgs
mechanism that explains how mass is acquired is only a
hypothetical event with no physical observations for its
verification available, the given physical property of a massive
particle should be the same after the Higgs mechanism and thus the
fermion should be redefined with mass in order to be consistent.
Otherwise, the fundamentality of the Dirac equation and mass would
be contradicted since the helicity is obtained from the massless
Dirac equation while other fundamental properties such as the
definition of antiparticle from the massive Dirac equation.
Therefore, the assumption of a massless fermion is invalid to
describe a massive fermion after the spontaneous symmetry breaking
and thus the Lorentz-violating approximation is still required for
the unification of electroweak interactions.

\section{Conclusion}
The massless approximation of the standard model proved to violate
Lorentz invariance as the exact fermion field in the rest frame
cannot be obtained from Lorentz transformations on the
approximated one. The assumption of massless fermion with the
Higgs mechanism fails to eliminate the necessity of this
approximation since the fermion should be redefined as a massive
particle after its mass is acquired. In conclusion, the
unification of electroweak interaction in the standard model is
only valid with the assumption of Lorentz violation.


\begin{thebibliography}
{}

\bibitem{Weinberg} S.\ Weinberg, Phys.\ Rev.\ Lett.\ {bf 19}
(1967), 1264.

\bibitem{Salam} A.\ Salam, in {\it Elementary Particle Theory}
({\it Nobel Symposium No. 8}, ed.\ N.\ Svartholm, Stockholm, 1968)
367.

\bibitem{Glashow} S.\ Glashow, Nucl.\ Phys.\ {\bf 22}, 579 (1961).


\bibitem{Higgs} P.W.\ Higgs, Phys. Rev. {\bf 145} (1966) 1156.

\bibitem{Kibble} T.W.B.\ Kibble, Phys.\ Rev.\ {\bf 155} (1967) 1554.

\bibitem{PeskinSchroeder} Michael E. Peskin and Daniel V.
Schoreder, {\it An introduction to Quantum Field Theory},
Addison-Wesley Publishing Co., (1995)


\bibitem{FeynmanGellMann} R.P.\ Feynman and M.\ Gell-Mann, Phys. Rev.
{\bf 109} (1958) 193.

\bibitem{Sudarshan} E.C.G.\ Sudarshan and R.E.\ Marshak, Phys.
Rev. {\bf 109} (1958) 1860.


\end{thebibliography}
\end{document}